\documentclass[pra,aps,twocolumn]{revtex4-1}
\usepackage{amsmath}
\usepackage{amssymb}
\usepackage{graphicx}

\newcommand{\mP}{{\mathcal P}}
\newcommand{\mT}{{\mathcal T}}

\newcommand{\beqa}{\begin{eqnarray}}
\newcommand{\eeqa}{\end{eqnarray}}

\begin{document}
\title{Dynamics, disorder effects, and $\mP\mT$-symmetry breaking in waveguide lattices with localized eigenstates} 
\author{Harsha Vemuri}
\email{These authors contributed equally to the project.}
\author{Vaibhav Vavilala}
\email{These authors contributed equally to the project.}
\author{Theja Bhamidipati}
\email{These authors contributed equally to the project.}
\author{Yogesh N. Joglekar}
\email{yojoglek@iupui.edu}
\affiliation{Department of Physics, 
Indiana University Purdue University Indianapolis (IUPUI), 
Indianapolis, Indiana 46202, USA}
\date{\today}
\begin{abstract} 
Recently, waveguide lattices with non-uniform tunneling amplitudes have been explored due to their myriad tunable properties, many of which arise from the extended nature of their eigenstates in the absence of disorder. Here, we investigate the dynamics, localization, and parity- and time-reversal-($\mP\mT$) symmetry breaking in lattices that support only localized eigenstates in the disorder-free limit. We propose three families of tunneling profiles that lead to qualitatively different single-particle time evolution, and show that the effects of weak disorder contain signatures of the localized or extended nature of clean-lattice eigenstates. We show that in lattices with localized eigenstates, the signatures of $\mP\mT$-symmetry breaking are acutely sensitive to the initial form of the wave packet. Our results suggest that waveguide lattices with localized eigenstates will exhibit a wide array of phenomena that are absent in traditional systems.  
\end{abstract}
\maketitle

\section{Introduction}
\label{sec:intro}

Over the past decade single-mode, coupled, optical waveguides~\cite{yariv} have become a new paradigm for the realization of an ideal, one-dimensional, tight-binding lattice model~\cite{review} with roughly constant tunneling amplitudes and on-site potentials~\cite{berg1,berg2}, as well as non-Hermitian parity- and time-reversal- ($\mP\mT$-) symmetric gain and loss potentials~\cite{ruter}. The electric field $E(k,z)$ in waveguide $k$ at a distance $z$ along the waveguide  obeys equation of motion that is identical to that of a time-dependent wave function $\psi(k,t)$; here $k$ is the lattice-site index and $t=z/v$ where $v=c/n_R$ is the speed of electromagnetic waves in the waveguide and $n_R$ is its refractive index. Optical waveguides have been used to simulate several phenomena such as the Anderson localization~\cite{berg1,berg12,berg2}, spontaneous $\mP\mT$-symmetry breaking~\cite{ruter}, Bloch oscillations~\cite{bloch2}, quantum random walks~\cite{qrw1,qrw2}, and Hanbury-Brown-Twiss correlations due to random tunneling and on-site potential~\cite{berg3}. Most recently, one-dimensional, tight-binding lattice models with a position-dependent tunneling have been explored. The energy spectrum, density of states, and single-particle wave packet evolution in these tunable waveguides can be varied over a wide range by choosing an appropriate tunneling function~\cite{avadh,clint}; in particular, tunneling functions that lead to commensurate energy levels, and the attendant wave packet reconstruction, have been extensively explored~\cite{gf,longhi}. 

All of these cases have focused on lattice models that, in the continuum limit, describe the dynamics of a non-relativistic, quantum particle with a position-dependent mass that is determined by the tunneling function. The eigenstates of the corresponding Hamiltonians are, thus, extended~\cite{avadh}. Therefore, the effects of diagonal (on-site) and off-diagonal (tunneling) disorder in such systems, and their signatures in Anderson localization~\cite{anderson} and the Hanbury-Brown-Twiss correlations~\cite{hbt}, correspond to those in condensed matter systems. Since these systems are typically probed through transport, traditionally, lattice models that have only localized eigenstates in the clean limit have not been studied; such models would result in a vanishing electrical conductivity even in the disorder-free limit. Optical waveguides, on the other hand, are probed via light intensity measurements along the waveguide, not by the motion of a wave packet across them. They offer an unprecedented ability to tailor the site-dependent tunneling~\cite{gf,nolte}; something that is difficult to do in condensed matter systems and, to a lesser degree, in optical lattices~\cite{bloch,zoller}.

In this paper, we investigate the dynamics, disorder effects and $\mP\mT$-symmetry breaking in non-uniform, tight-binding lattice models in which most of the eigenstates are localized in {\it the absence of disorder}. We propose three classes of tunneling functions that demonstrate the wide range of properties typical of such models. Our three primary results are as follows: (i) The energy spectra and wave packet time-evolution have properties with no counterparts in systems with extended eigenstates. (ii) The wave packet localization due to a weak disorder encodes the (extended or localized) nature of eigenstates even for lattices with identical clean-limit energy spectra. (iii) Singatures of $\mP\mT$-symmetry breaking are acutely sensitive to the initial form of the wave packet. Our results show that light propagation can be significantly controlled and manipulated in coupled waveguides with localized eigenstates. 

The plan of the paper is as follows. In the following paragraphs, we introduce the tight binding model for a non-uniform $N$-site lattice. Section~\ref{sec:cl} presents the properties of a clean lattice for the three classes of tunneling functions that we introduce. We consider the effects of a weak disorder on the time-evolution in such lattices in Sec.~\ref{sec:disorder}. In Sec.~\ref{sec:pt}, we expand these models to their $\mP\mT$-symmetric counterparts, and consider the effect of a single pair of balanced loss and gain impurities at mirror-symmetric positions. We conclude the paper with a brief discussion in Sec.~\ref{sec:disc}. 

A lattice of $N$ coupled waveguides is described by the following Hamiltonian, 
\begin{equation}
\label{eq:tb}
H=-\sum_{i=1}^{N-1}t(i) \left( |i+1\rangle\langle i |+ |i\rangle\langle i+1|\right)+\sum_{i=1}^Nv_i |i\rangle\langle i|,
\end{equation}
where $|k\rangle$ represents the state with a (single) particle at site $k$, $t(k)>0$ is tunneling amplitude between sites $k$ and $k+1$, and $v_k$ is the on-site potential determined by the local index of refraction at site $k$. Hamiltonian~(\ref{eq:tb}) represents open boundary conditions, $t(0)=0=t(N)$. Note that although $t(k), v_k$ have the units of energy, we will use their scaled versions, $t(k)/\hbar c, v_k/\hbar c$, which have the units of inverse-length, in the discussion of possible sample parameters~\cite{berg12,qrw2,nolte}. 

We choose tunneling amplitude profiles characterized by two continuous, distinct, functions. $t_O(2k-1)$ on the odd sites and $t_E(2k)$ on the even sites are chosen such that $t_E(2k)/t_O(2k-1)\ll 1$ for most $k\sim O(N)$. In the limit $t_E\equiv 0$, the system decouples into $N/2$ waveguide-pairs or dimers, the energy spectrum of Eq.(\ref{eq:tb}) is given by $\mp t_O(1),\mp t_O(3),\ldots$, and the eigenfunctions are localized, symmetric ($S$) and antisymmetric ($A$), dimer states given by $|S(A),2k-1\rangle=(|2k-1\rangle\pm|2k\rangle)/\sqrt{2}$. Thus, the particle-hole symmetric energy spectrum of the clean-lattice Hamiltonian is determined by the odd-tunneling function $t_O$. 

When $t_E\neq 0$, the Hamiltonian in the dimer basis becomes a block-tridiagonal, symmetric matrix with the diagonal block $D_{2k-1}$ and off-diagonal block $T_{2k-1,2k+1}=T^\dagger_{2k+1,2k-1}$ given by 
\begin{eqnarray}
\label{eq:d}
D_{2k-1}& = &  t_O(2k-1)\mathrm{diag}\left[-1,+ 1\right],\\ 
\label{eq:t}
T_{2k-1,2k+1} & = & t_E(2k)\left[
\begin{array}{cc}
+1 & +1\\
-1 & -1\\
\end{array}
\right].
\end{eqnarray}
In the following section, we explore the consequences of such tunneling profiles for a disorder-free lattice. 

\section{Dynamics in a clean lattice} 
\label{sec:cl}

We start with three tunneling functions, 
\begin{eqnarray}
\label{eq:power}
t_p(k) & = & \left\{
\begin{array}{cc}
k^2 & \mathrm{odd}, \\
k & \mathrm{even},\\
\end{array}
\right.\\
\label{eq:log}
t_l(k) & = & \left\{
\begin{array}{cc}
k & \mathrm{odd},\\
\beta\ln(k) & \mathrm{even},\\
\end{array}
\right.\\
\label{eq:ext}
t_e(k) & = & \left\{
\begin{array}{cc}
\left[k(N-k)\right]^{1/2} & \mathrm{odd},\\
\beta\ln[k(N-k)] & \mathrm{even,}\\
\end{array}
\right.
\end{eqnarray}
each of which satisfies the criterion $t_E/t_O\ll 1$ and the energy-scale prefactor $t'$ in Eqs.~(\ref{eq:power})-(\ref{eq:ext}) has been set to unity, $t'=1$; recall that for traditional waveguide lattices with constant tunneling, $t'/\hbar c\sim 10^2-10^4$ m$^{-1}$~\cite{berg12,qrw2,nolte}. As we will show below, {\it the functional forms} - the power-law tunneling function $t_p(k)$, logarithmic tunneling function $t_l(k)$, and the tunneling function $t_e(k)$ that supports extended states - {\it of the tunneling amplitude} broadly dictate the results that follow. 

\begin{figure}[h!]
\begin{center}
\begin{minipage}{1.05\columnwidth}
\begin{tabular}{cc}
\hspace{-5mm}
\begin{minipage}{0.5\columnwidth}
\includegraphics[angle=0,width=\columnwidth]{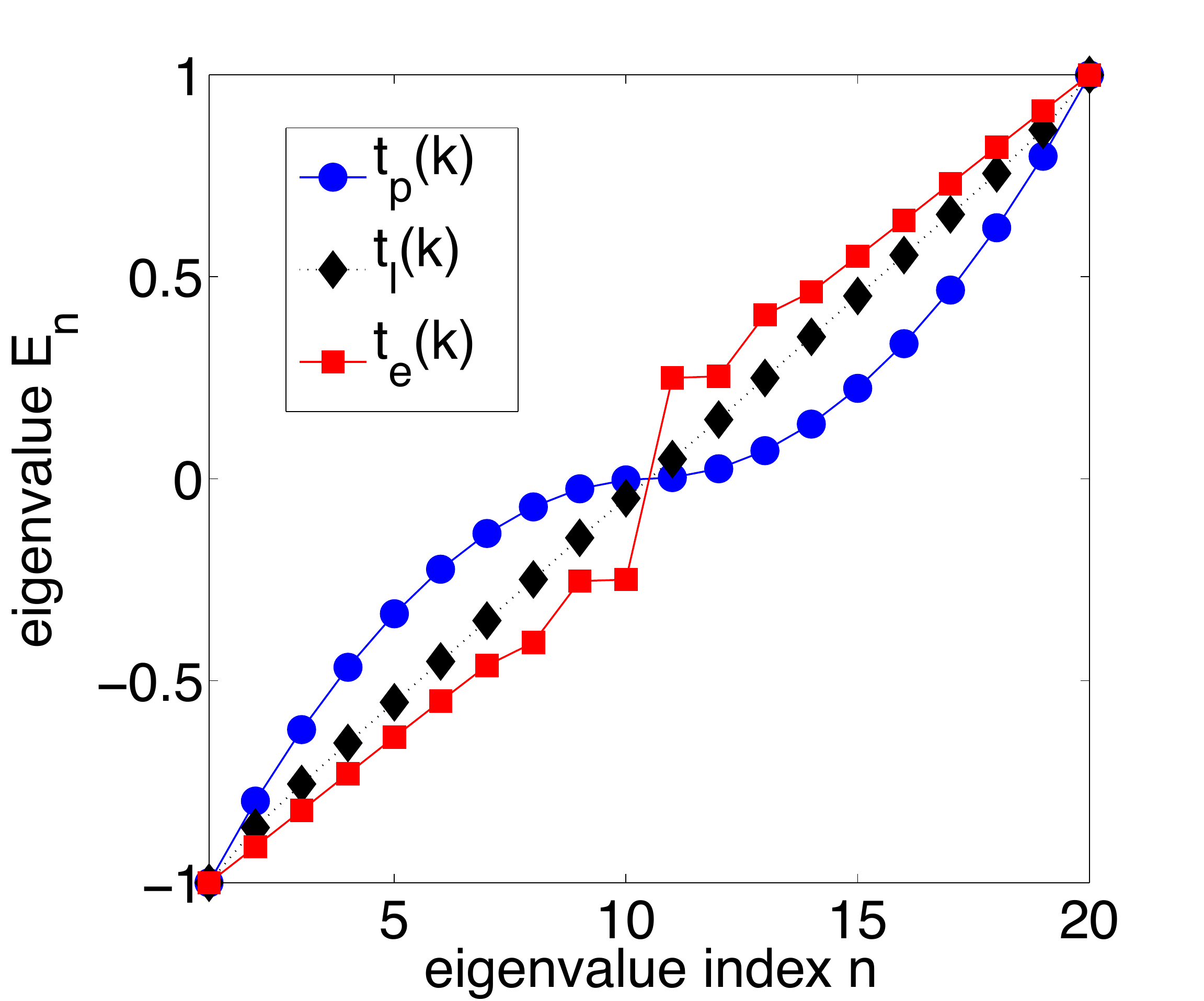}
\end{minipage}&
\hspace{-5mm}
\begin{minipage}{0.5\columnwidth}
\includegraphics[angle=0,width=\columnwidth]{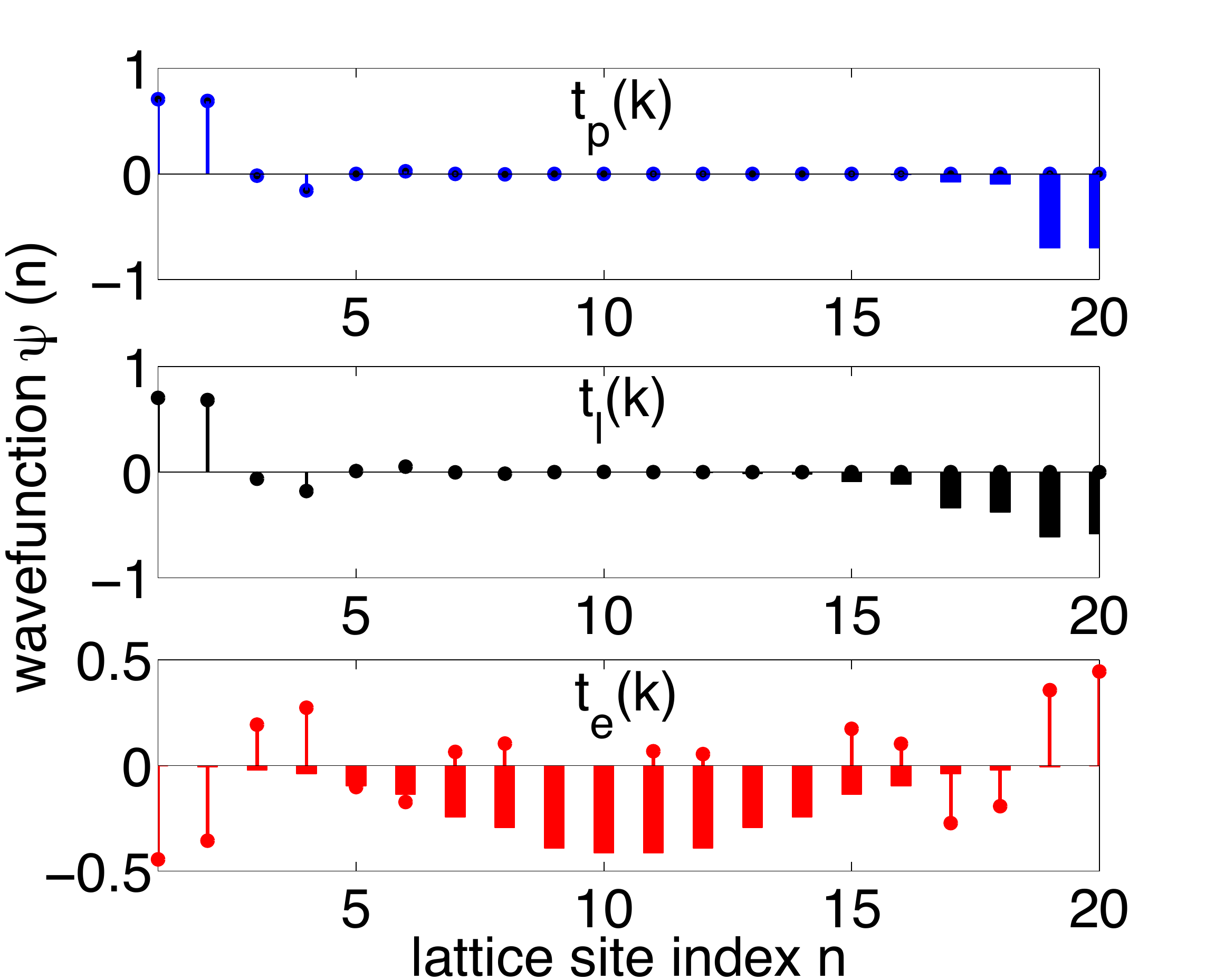}
\end{minipage}\\
\end{tabular}
\end{minipage}
\caption{(color online) a) Left: Typical energy spectra for lattices with tunneling functions $t_p(k)$ (blue circles), $t_l(k)$ (black diamonds), and $t_e(k)$ (red squares), and $N=20, \beta=1$; the energy eigenvalues are scaled by their maximum. For $t_p(k)$ (blue circles) the spectrum is quadratic, for $t_l(k)$ (black diamonds) it is mostly linear, and for $t_e(k)$ it is linear near the band edges. The primary features of the spectra are determined by the odd-tunneling function $t_O(2k-1)\gg t_E(2k)$. b) Right: Dimensionless eigenfunctions for ground-state (bars) and the center-band state (stems). The top panel shows ground state localized at the last dimer and the center-band state localized at the first dimer. The center panel shows a broadened ground state, localized near the last dimer, and the center-band state localized at the first dimer. The bottom panel shows an extended, Gaussian, ground state centered at $N/2$ and the center-band state localized near both edges.}
\label{fig:energy}
\end{center}
\vspace{-5mm}
\end{figure}
The left-hand panel in Fig.~\ref{fig:energy} shows typical, particle-hole symmetric~\cite{lattice}, energy spectra for a lattice with $N=20,\beta=1$ for $t_p(k)$ (blue circles), $t_l(k)$ (black diamonds), and $t_e(k)$ (red squares). We use the maximum energy $E_{\max}=-E_{\min}$ to define the energy and time scales; it varies as $E_{p,\max}(N)\sim t' N^2$, $E_{l,\max}(N)\sim t' N\sim E_{e,\max}(N)$ for $N\gg 1$. Note that the energy spectrum for $t_l(k)$ is {\it exactly linear} when $\beta=0$, and a nonzero $\beta\ll N$ creates deviations from linearity at the band edges~\cite{caveat}. On the other hand, the linearity of the spectrum for $t_e(k)$ arises from the exactly linear spectrum of $T(k)=[k(N-k)]^{1/2}$~\cite{avadh,gf,longhi}, and the spectrum become nonlinear when $\beta\ll 1$. Note that these results are true for any $N\gg 1$. Given the wide range over which the tunneling amplitude~\cite{qrw2,nolte} and the number of waveguides in a lattice can be varied ($N\sim 10-100$)~\cite{berg1,berg12,qrw2}, the construction of waveguide lattices with tunneling profiles (\ref{eq:power})-(\ref{eq:ext}) seems feasible.

It follows from earlier analysis that the ground state of Hamiltonian (\ref{eq:tb}) is localized near the dimer with largest internal tunneling, $\max t_O$, whereas the eigenstate near the center of the band is localized near the dimer with smallest internal tunneling, $\min t_O$; this is true regardless of the number of lattice sites $N$. The right-hand panels in Fig.~\ref{fig:energy} show these two eigenfunctions. The top panel, with a power-law tunneling $t_p(k)$, shows the ground-state (blue bars) and the center-band state (blue stems). The center panel, with a log tunneling $t_l(k)$, shows a broad ground state (black bars) and narrow center-band state (black stems); this difference between the ground-state sizes is because the ratio $t_E(N-1)/t_O(N)$, a measure of perturbation away from the dimer picture, is smaller for $t_p(k)$ than it is for $t_l(k)$. The bottom panel shows the ground state (red bars) and center-band state (red stems) for $t_e(k)$; we remind the reader that the ground state for the tunneling function $T(k)$ is a Gaussian centered at site $N/2$~\cite{avadh}. Results in Fig.~\ref{fig:energy} suggest that in spite of the perturbation introduced by the even-site hopping functions $t_E$ in Eqs.~(\ref{eq:power}) and (\ref{eq:log}), the localized nature of eigenstates is preserved.  

\begin{figure}[h!]
\begin{center}
\includegraphics[angle=0,width=0.9\columnwidth]{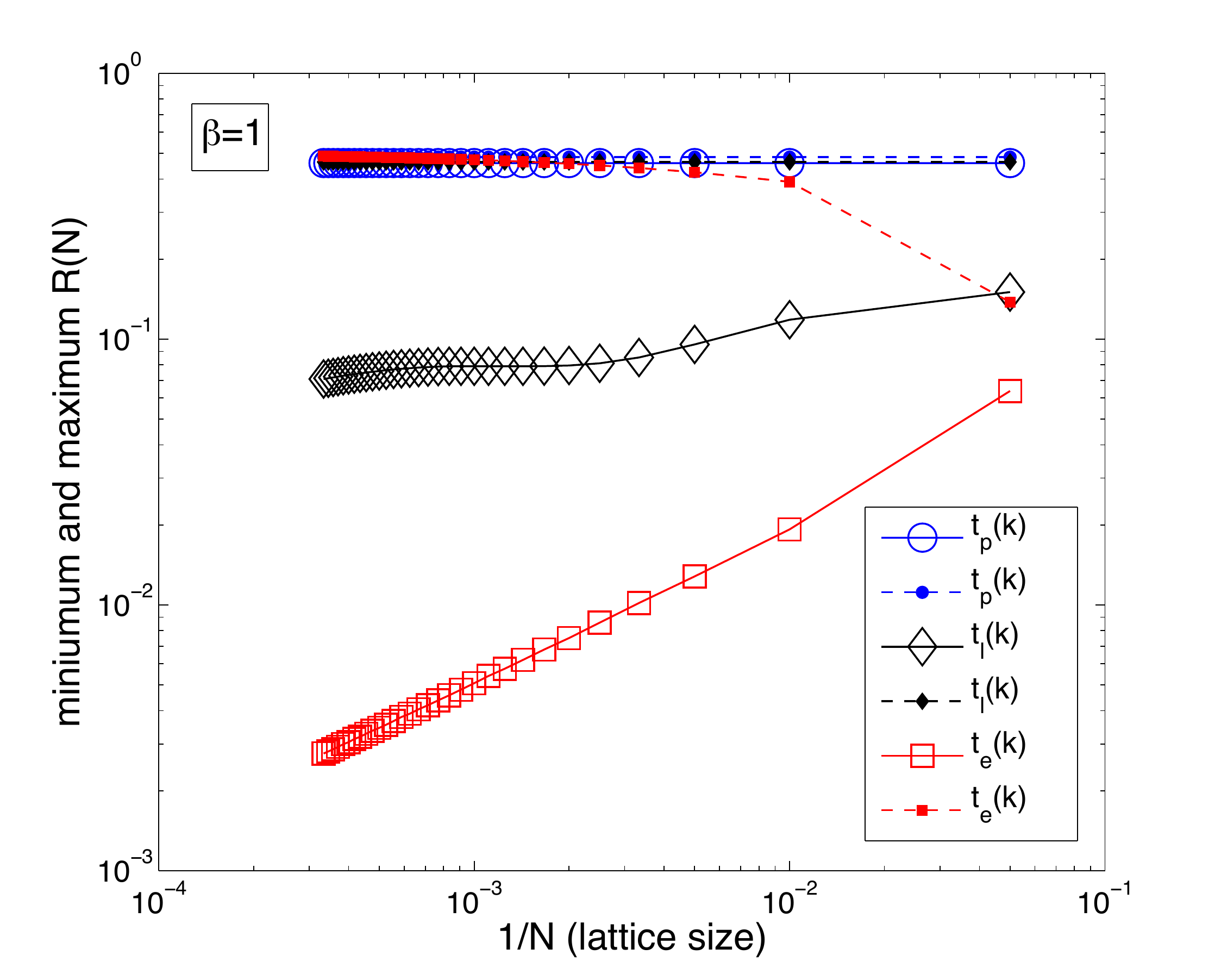}
\caption{(color online) Minimum (open symbols) and maximum (filled symbols) inverse participation ratios $R(N)$ as a function of lattice size $N$ for the three tunneling functions with $ \beta=1$; note the logarithmic scale on both axes. For power-law tunneling $t_p(k)$ (blue), all states are localized, as are those for the log-tunneling function $t_l(k)$ (black). The tunneling function $t_e(k)$ (red squares) has both localized and extended states. When $\beta=0$, all eigenstates - dimers - are localized, and as $\beta$ increases, the minimum $R$ values for $t_l(k)$ and $t_e(k)$ decrease.}
\label{fig:ipr}
\end{center}
\vspace{-5mm}
\end{figure}
To quantify this claim, we calculate the $N$-dependence of inverse participation ratio $R(N)$ for all eigenstates. The inverse participation ratio $R_\psi(N)$ for a normalized eigenstate $|\psi\rangle=\sum_{i=1}^N f_i |i\rangle$ is defined as
\begin{equation}
\label{eq:ipr}
R_\psi(N)=\sum_{i=1}^N |f_i|^4.
\end{equation}
As $N\rightarrow\infty$, $R_\psi(N)\leq 1$ saturates to a nonzero value for a localized state and vanishes, $R_\psi(N)\sim N^{-\alpha}$ with $\alpha>0$, for an extended state~\cite{ipr}. Figure~\ref{fig:ipr} shows the minimum and maximum values of $R(N)$ for lattices with $N=20-3000$ and the three tunneling functions, Eqs.~(\ref{eq:power})-(\ref{eq:ext}), with $\beta=1$; note the logarithmic scale on both axes. For power-law tunneling $t_p(k)$, the minimum (blue open circles) and maximum (blue filled circles) values of $R(N)$ are nonzero, almost equal to each other, and indicate that all localized eigenstates have approximately the same size. The nonzero minimum (black open diamonds) and maximum (black filled diamonds) $R(N)$ values for the log-tunneling function $t_l(k)$ imply that the ground state, although localized, is broader than the center-band state (see Fig.~\ref{fig:energy}). For tunneling function $t_e(k)$, inverse participation ratios show the existence of both localized (red filled squares) and extended (red open squares) states. These results show that most, if not all, states of a {\it disorder-free} lattice with these tunneling profiles are localized.  

\begin{figure}[h!]
\begin{center}
\includegraphics[angle=0,width=1.05\columnwidth]{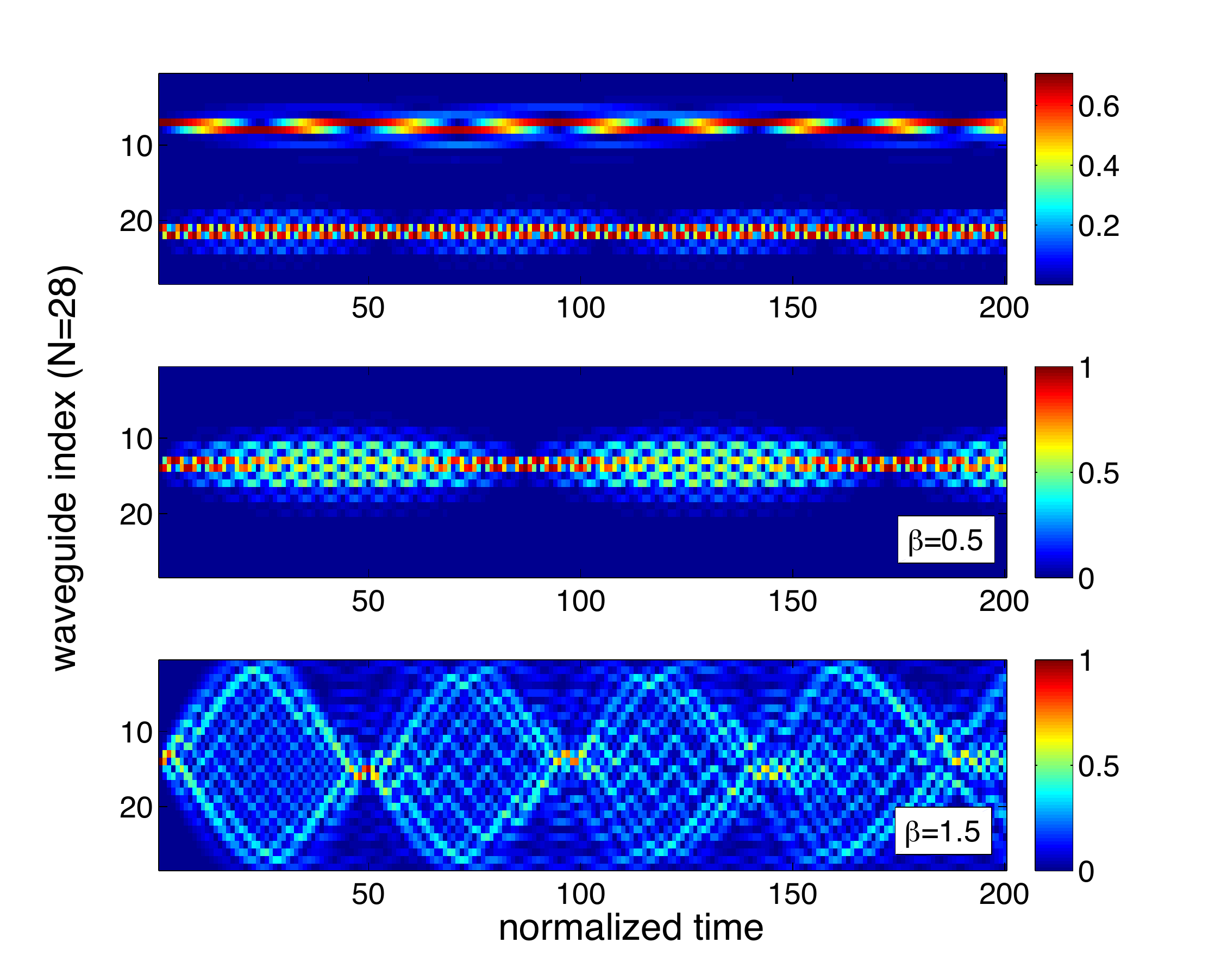}
\caption{(Color online) Time-dependent amplitude ${\mathcal A}(k,t)$ for lattices with $N=28$. Top panel: power-law tunneling function $t_p(k)$ leads to Rabi oscillations with waveguide-index dependent frequency determined by the local gap $2t_O$. Center panel: log-tunneling function $t_l(k)$ with $\beta=0.5$ leads to periodic wave packet reconstruction due to its linear spectrum. Note that increasing $\beta$ leads to asymmetrical increase in the vertical spread of the wave packet, as is expected due to the higher even-tunneling amplitude at larger waveguide index. Bottom panel: tunneling function $t_e(k)$ with $\beta=1.5$ has a partially linear spectrum and 
some extended states. For a fixed $N$, the maximum wave packet spread shown in the center and bottom panels increases monotonically with $\beta$. The bottom panel shows that for a moderate value of $\beta=1.5\ll N=28$, the wave packet spreads over the entire array, even though for $\beta=0$ it is localized on the dimer. }
\label{fig:psi}
\end{center}
\vspace{-5mm}
\end{figure}

Now, we study the time-evolution of a wave packet that is initially confined to one (or two) waveguides. Figure~\ref{fig:psi} shows the amplitude ${\mathcal A}(k,t)= |\langle k|\psi(t)\rangle|$ of the time-evolved wave functions $|\psi(t)\rangle=\exp[-iHt/\hbar]|\psi(0)\rangle$ for the three tunneling functions in a disorder-free lattice with $N=28$ sites.  The horizontal axis in each panel denotes time normalized by $\hbar/E_{\max}$ for each tunneling function, $t/(\hbar/E_{\max})$. The corresponding distances along the waveguide, for a normalized time-range 
$t/(E_{\max}/\hbar)=100$, are given by $z=ct \sim (10-10^3)/N^2\sim 0.01-1$ mm for the power-law tunneling function $t_p(k)$ and $z=ct\sim (10-10^3)/N\sim 0.3-30$ mm for the other two tunneling functions. Note that waveguides with constant tunneling amplitude are a few mm long~\cite{berg12,qrw2}; thus, they correspond to a time-range $t/(E_{\max}/\hbar)\sim 10^4$ for power-law tunneling function and a time-range of $t/(E_{\max}/\hbar)\sim 100-1000$ for the log-tunneling function. This is consistent with the observation~\cite{clint} that for a non-uniform waveguide array with a given physical length can be used to explore short-time or long-time behavior based on its bandwidth. 

The top panel in Fig.~\ref{fig:psi} shows ${\mathcal A}(k,t)$ for tunneling function $t_p(k)$, and $|\psi(0)\rangle=(|N/4\rangle+|3N/4\rangle)/\sqrt{2}$. Since all states of this Hamiltonian are strongly and equally localized, the partial wave packets located on two spatially separated dimers evolve independently. Each undergoes Rabi oscillations with the local frequency, and the ratio $\sim 9=3^2$ of these frequencies is equal to the ratio of {\it A-S} energy gap $2 t_O$ for the dimers at $3N/4$ and $N/4$ respectively. The (blue) side-wings in each case indicate the mixing with adjacent dimers due to the small, but nonzero, even-tunneling function $t_E(2k)/ t_O(2k-1)\sim 1/k$. The center panel corresponds to $t_l(k)$ with $\beta=0.5$, and an initial state localized at the central waveguide, $|\psi(0)\rangle=|N/2\rangle$. Due to the linear spectrum, the wave packet is reconstructed after a time $T_l=2\pi\hbar/\Delta E_l$ where $\Delta E_l\approx 2E_{l,\max}/N$ is the (approximately) constant energy-level spacing (see Fig.~\ref{fig:energy}). The time-dependent width of the wave packet is due to the even-tunneling function $t_E/t_O\sim\beta\ln(k)/k>1/k$ and it increases with $\beta$ for $\beta\ll N$~\cite{caveat}. The bottom panel shows ${\mathcal A}(k,t)$ for the tunneling function $t_e(k)$, which has both localized and extended eigenstates, and $|\psi(0)\rangle=|N/2\rangle$. We see that, although the wave packet remains confined to the dimer when $\beta\ll 1$, it spreads across all waveguides for a moderate value of $\beta=1.5\ll N$. In addition, as expected, it undergoes partial reconstructions that are reminiscent of (and due to) the perfect revival that occurs in lattices with tunneling function $T(k)=[k(N-k)]^{1/2}$~\cite{clint,longhi}. 

Thus, we predict that {\it the dynamics of a wave packet in the disorder-free lattices can be systematically controlled} by a suitable value of $\beta$ and the initial waveguide location. 

\section{Effect of a weak disorder}
\label{sec:disorder}

For an infinite, one-dimensional system, an arbitrarily weak disorder exponentially localizes all states~\cite{anderson}. In a finite waveguide lattice, an arbitrarily weak disorder localizes the wave packet to its initial waveguide $m_0$. The localized fraction, characterized by the intensity at waveguide $m_0$, saturates with time (or distance along the waveguide), but increases with the disorder strength $v_0$. Therefore, the long-time, steady-state, disorder-averaged intensity $I(k)=|\langle k|\psi(t)\rangle|^2_{v_0}$ has a maximum at $k=m_0$ and, for $1\ll m_0\leq N/2$, it symmetrically decays with exponential tails away from $m_0$~\cite{berg1,anderson}. At short times, however, the time- and waveguide-dependent intensity $I(k,t)$ is determined by the competition between the spread dictated by the disorder-free lattice spectrum and localization due to the disorder. 

In the following, we focus on the steady-state intensity $I(k)$ for two lattices, with tunneling functions $T(k)$ and $t_l(k)$ respectively. Both of them have (nearly) identical, linear, energy spectra in the disorder-free limit. However, all eigenstates of the former are extended and the latter are localized in the absence of disorder. 
\begin{figure}[h!]
\begin{center}
\hspace{-6mm}
\includegraphics[angle=0,width=\columnwidth]{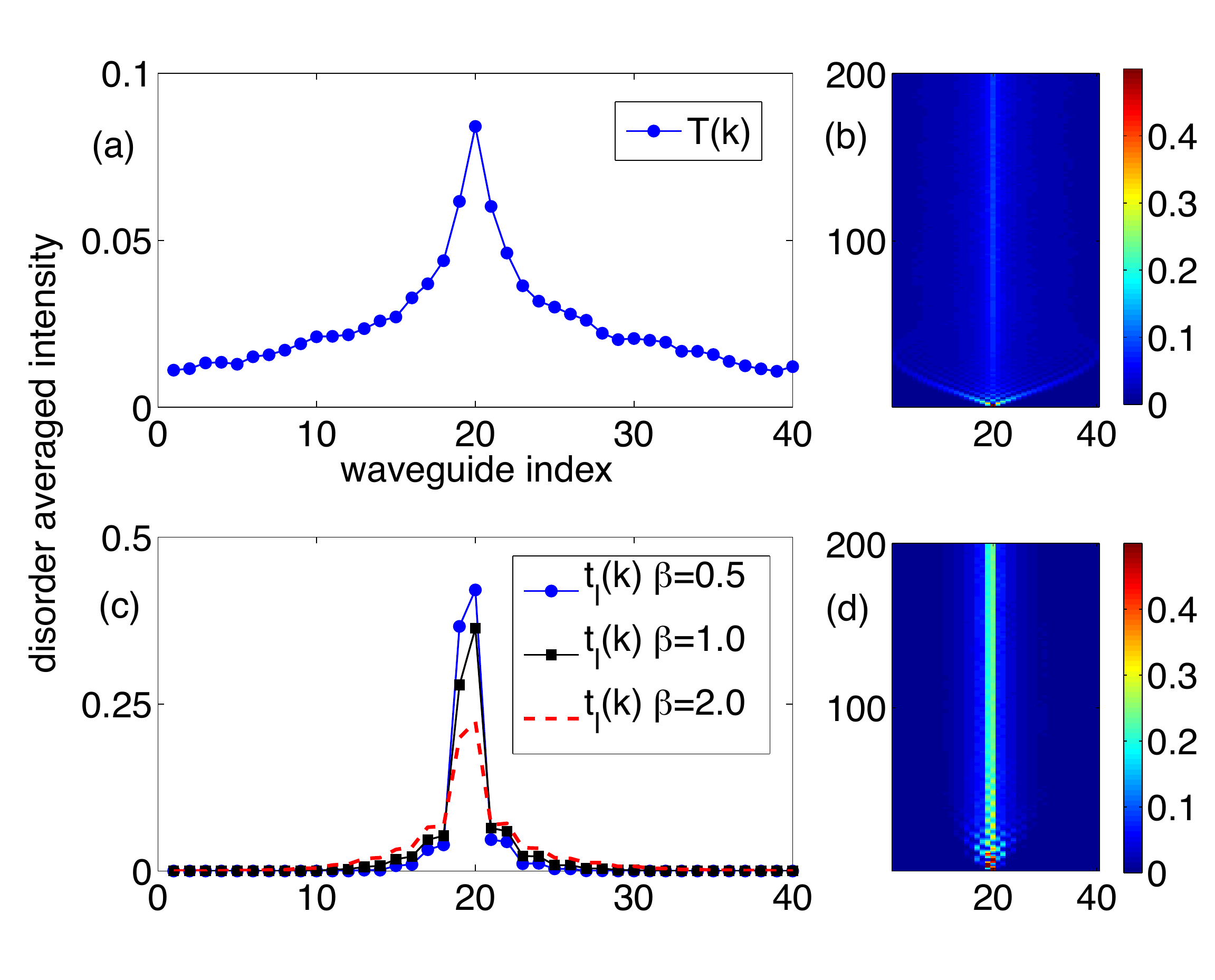}
\caption{(Color online) (a) Steady-state, disorder-averaged intensity in a waveguide lattice with $N=40$, tunneling function $T(k)$, and initial wave packet $|\psi(0)\rangle=|N/2\rangle$ shows localization with exponential tails. (b) The time-dependent intensity $I(k,t)$ for the same. The disorder strength is $v_0/E_{\max}=0.2$. (c) Same as panel (a), but with tunneling function $t_l(k)$ as a function of $\beta$. For small $\beta\leq 1$, the localized fraction is confined largely to the dimer $(N/2-1,N/2)$. With increasing $\beta$, however, the weight at neighboring waveguides increases, the weight at the central maximum decreases, and the intensity profile approaches that in panel (a); note the scale-difference in the two panels. (d) The time-dependent intensity $I(k,t)$ for tunneling function $t_l(k)$ with $\beta=2$ shows a larger fraction of the wave packet localized at the initial waveguide compared to its value in panel (b). }
\label{fig:disorder}
\end{center}
\vspace{-5mm}
\end{figure}

Figure~\ref{fig:disorder} shows results for a weak disorder, $v_0/E_{\max}=0.2$, in lattices with $N=40$ sites, and tunneling functions $T(k)$ (top row) and $t_l(k)$ (bottom row). We have verified that the results are independent of the number of realizations $N_r\sim 10^3$ for the disorder potential $v_k$, and the type of disorder-potential distribution (Gaussian, uniform) as long the distribution as zero mean and variance $v_0$. Panels (b) and (d) show typical time-dependent intensity $I(k,t)$ for a state $|\psi(0)\rangle=|N/2\rangle$; horizontal (vertical) axis is the waveguide index (normalized time). Recall that in the absence of disorder, $v_0=0$, the intensity time-evolution in the two panels should be similar to that in Fig.~\ref{fig:psi}. When disorder is included, the localized fraction for tunneling function $t_l(k)$, with $\beta=2$, is greater than that for $T(k)$. Panels (a) and (c) show corresponding steady-state intensity $I(k)$. Panel (a) shows that the wave packet is localized with exponential tails, as is expected for a lattice with extended eigenfunctions in the clean limit~\cite{berg1,avadh}. Panel (b) shows qualitatively different intensity profiles as a function of $\beta$; note the vertical-scale difference in panels (a) and (b). As $\beta$ increases from $0.5$ (blue circles) to $2.0$ (red dashed line), the intensity maximum at the center reduces, {\it while the intensity at the neighboring waveguides increases}.  Eventually, the intensity profile approaches that in panel (a). 

These results demonstrate the crucial role played by qualitatively different eigenstates of (approximately) isospectral Hamiltonians in the presence of disorder, and the change that the localization profile undergoes when going from one to the other. For Hamiltonians with tunneling functions~(\ref{eq:power})-(\ref{eq:ext}), the continuum limit is a massive particle with an internal, dimer, degree of freedom. Therefore, its disorder localization profile is different. 

\section{$\mP\mT$-symmetry Breaking} 
\label{sec:pt}

In this section, we will explore disorder-free lattices with localized eigenstates in the presence of a single pair of $\mP\mT$-symmetric, gain and loss impurities $\pm i\gamma$~\cite{ruter}. To this end, we consider a lattice with $2N$ waveguides and define the $\mP\mT$-symmetric extension of the tunneling function, 
\begin{eqnarray}
\label{eq:pt}
t_{\mP\mT}(k) & = & \left\{
\begin{array}{cc}
t(k) & 1\leq k<N, \\
t_C & k=N,\\
t(2N-k) & N+1\leq k<2N,\\
\end{array}
\right.
\end{eqnarray}
where $t(k)$ is either a power-law or log tunneling function, Eqs.~(\ref{eq:power})-(\ref{eq:log}), and $t_{\mP\mT}(N)=t_C$ is arbitrary. Note that it is straightforward to modify this definition when the number of waveguides in an array is odd. The non-Hermitian, $\mP\mT$-symmetric Hamiltonian for the system is then given by
\begin{equation}
\label{eq:Hpt}
H_{\mP\mT}= H( t_{\mP\mT}) +i\gamma ( |m\rangle\langle m|-|\bar{m}\rangle\langle\bar{m}|)\neq H^{\dagger}_{\mP\mT},
\end{equation}
where $H(t_{\mP\mT})$ is Hamiltonian~(\ref{eq:tb}) with the tunneling function (\ref{eq:pt}) or $T(k)=[k(2N-k)]^{1/2}=T(2N-k)$, $m$ is the position of the gain impurity, and $\bar{m}=2N+1-m$ denotes its mirror position where the loss impurity is located. Although $H_{\mP\mT}$ is not Hermitian, all of its eigenvalues are real when $\gamma<\gamma_c(m)$. The critical impurity strength $\gamma_c(m)$, in general, is a function of the tunneling profile and the distance $d=|m-\bar{m}|$ between the loss and gain impurities. However, in the special case of nearest neighbor impurities, $m=N$, the critical impurity strength is given by $\gamma_c(N)=t_C$ and all energy eigenvalues simultaneously become complex when $\gamma>\gamma_c(N)=t_C$~\cite{jake}. We choose $t_C=t_{\mP\mT}(N-1)=t_{\mP\mT}(N+1)$, so that the tunneling amplitude at the lattice center is continuous. In the following, we focus on the evolution of time-dependent intensity across the $\mP\mT$-symmetry threshold, $\gamma/\gamma_c=1.00\mp 0.01$. Note that since the Hamiltonian $H_{\mP\mT}$, Eq.(\ref{eq:Hpt}),  is not Hermitian, the time evolution operator $\exp[-iH_{\mP\mT}t/\hbar]$ is not unitary and the total intensity $I(t)=\sum_{k=1}^{2N} I(k,t)$ is not conserved. As in Sec.~\ref{sec:disorder}, we consider two tunneling functions: the first, $\mP\mT$-symmetric version of the log-tunneling function, Eq.(\ref{eq:log}), with $\beta=0.5$, has a linear spectrum and purely localized eigenstates; the second, $T(k)=[k(2N-k)]^{1/2}$ for a lattice with $2N$ sites, has a linear spectrum and only extended eigenstates. 

\begin{figure}[htb]
\begin{center}
\hspace{-6mm}
\includegraphics[angle=0,width=1.05\columnwidth]{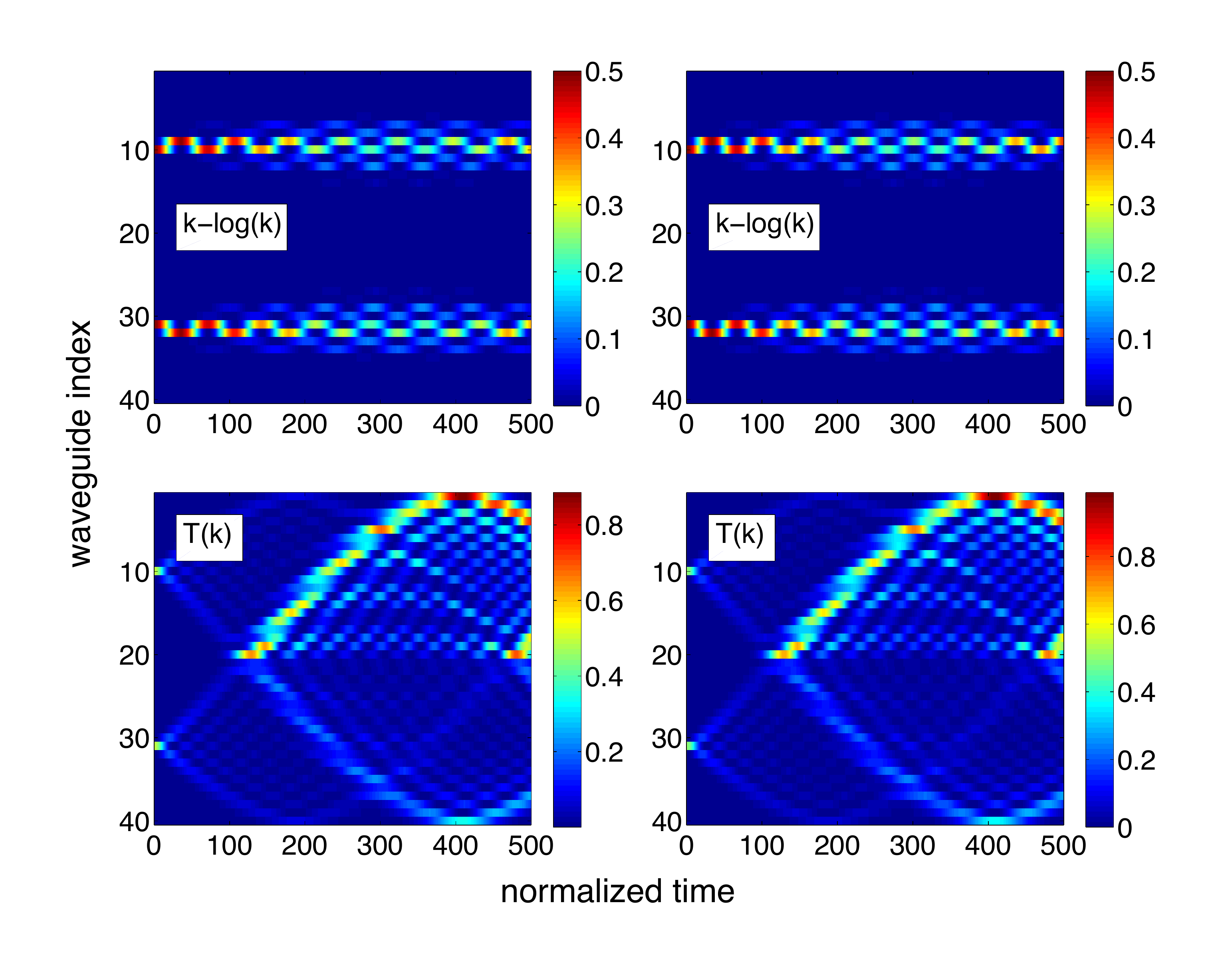}
\caption{(Color online) Intensity $I(k,t)$ for log (top row) and $T(k)$ (bottom row) tunneling functions in a lattice with $2N=40$ sites, gain and loss impurities $\pm i\gamma$ at positions $(20,21)$, and initial state $|\psi(0)\rangle_f=(|10\rangle+|31\rangle)/\sqrt{2}$ localized far away from them. The impurity strength is below the threshold, $\gamma/\gamma_c=0.99$, for the left-hand column and above the threshold, $\gamma/\gamma_c=1.01$, for the right-hand column. The top row shows that $I(k,t)$ does not change significantly across the threshold, and the maximum intensity does not change from its Hermitian limit. The bottom row shows that the maximum intensity is higher than its $\gamma=0$ limit since the wave packet comes across the impurities, and there is a small increase in the intensity as the threshold is passed.}
\label{fig:far}
\end{center}
\vspace{-5mm}
\end{figure}
Figure~\ref{fig:far} shows the intensity $I(k,t)$ for a particle away from the impurities, with an initial state $|\psi(0)\rangle_f=(|N/2\rangle+ |3N/2+1\rangle)/\sqrt{2}$ in lattice with $2N=40$ sites and nearest neighbor impurities at $(m,\bar{m})=(20,21)$.  The left-hand (right-hand) column corresponds to impurity strength $\gamma/\gamma_c$ below (above) the $\mP\mT$-symmetry breaking threshold. The top row shows that the time-dependent intensity $I(k,t)$ for the $\mP\mT$-symmetric log-tunneling function does not change appreciably as impurity strength increases from $\gamma/\gamma_c=0.99$ (left-hand panel) and $\gamma/\gamma_c=1.01$ (right-hand panel); in addition, the maximum intensity is the same as its corresponding value in the Hermitian limit. These results are expected since the initial wave packet, localized away from the lattice center, does not ``come across" the gain and loss impurities. The bottom row shows corresponding results for tunneling function $T(k)$. We see that the intensity profile does not change significantly, although there is a minor enhancement for $\gamma/\gamma_c=1.01$ (right-hand panel) compared to $\gamma/\gamma_c=0.99$ (left-hand panel), and the maximum intensity is higher than its $\gamma=0$ value. These results are due to the extended nature of eigenstates, which ensures that any wave packet ``comes across'' the gain and loss impurities. 

\begin{figure}[htb]
\begin{center}
\hspace{-6mm}
\includegraphics[angle=0,width=1.05\columnwidth]{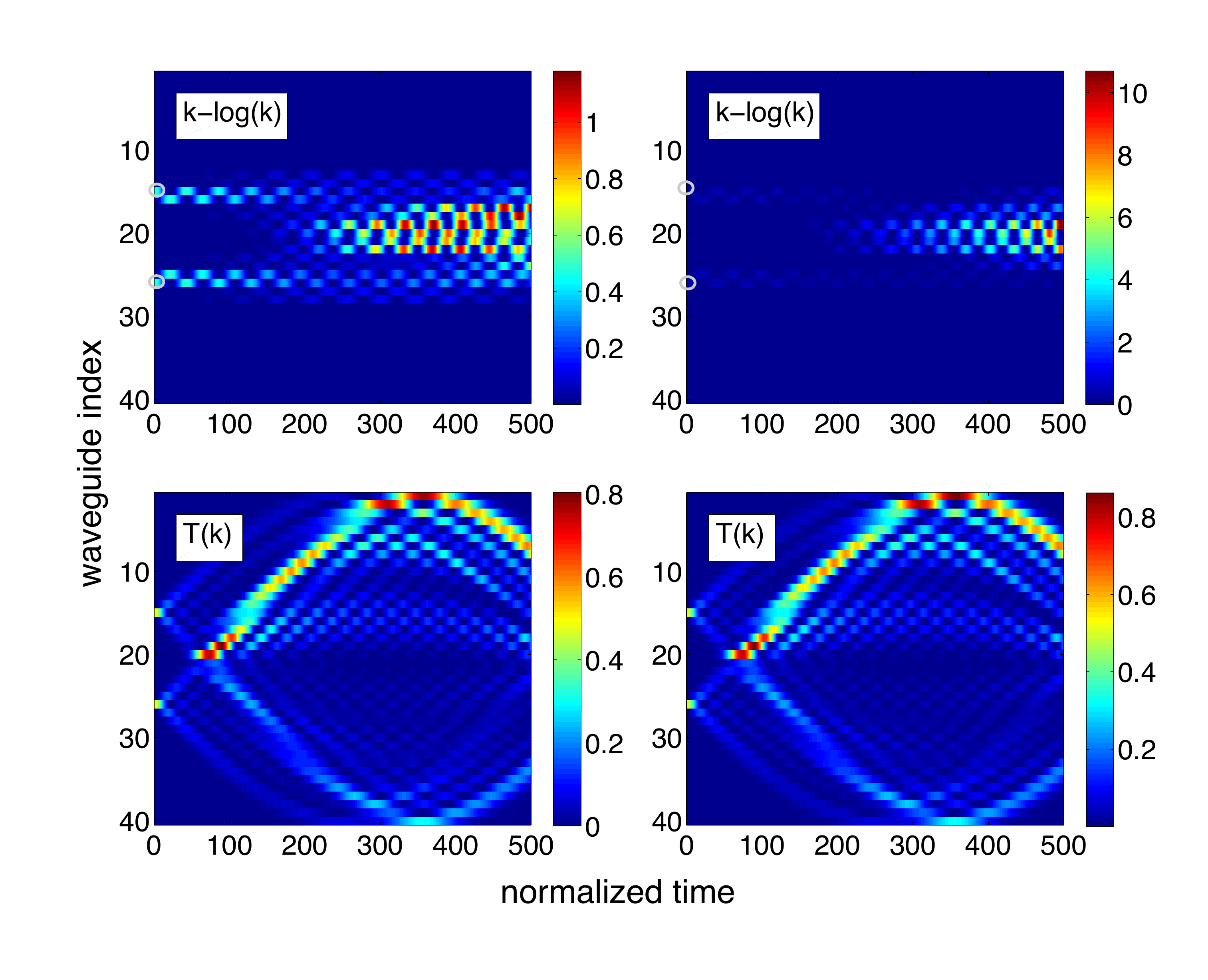}
\caption{(Color online) Intensity $I(k,t)$ for log (top row) and $T(k)$ (bottom row) tunneling functions in a lattice with $2N=40$ sites, gain and loss impurities $\pm i\gamma$ at positions $(20,21)$, and initial state $|\psi(0)\rangle_c=(|15\rangle+|26\rangle)/\sqrt{2}$ localized close to them. The impurity strength is below the threshold, $\gamma/\gamma_c=0.99$, for the left-hand column and above the threshold, $\gamma/\gamma_c=1.01$, for the right-hand column. The bottom row shows that that the maximum intensity is higher than its $\gamma=0$ limit since the wave packet comes across the gain and loss impurities, and there is a small increase in the intensity as the threshold is passed; see the bottom row, Fig.~\ref{fig:far}. The top row shows that the intensity increases by an order of magnitude as the impurity strength $\gamma$ is varied from below the threshold (left-hand panel) to above the threshold (right-hand panel); see the top row, Fig.~\ref{fig:far}.}
\label{fig:close}
\end{center}
\vspace{-5mm}
\end{figure}
How do these results change when the initial state is closer to the impurities? Figure~\ref{fig:close} shows corresponding results for the same lattice, same impurity strengths, and the same time-range, but with initial state $|\psi(0)\rangle_c=(|3N/4\rangle + |5N/4+1\rangle)/\sqrt{2}$. The bottom row shows that the intensity profile does not change significantly from the left-hand  panel, $\gamma/\gamma_c=0.99$ to the right-hand panel, $\gamma/\gamma_c=1.01$, although there is a minor enhancement in the right-hand panel (see the bottom row, Fig.~\ref{fig:far}). This is because the weight of any extended eigenstate at the gain-impurity site is small, and therefore, the effect of encountering the gain-site on a wave packet is small. The top row shows corresponding results for the log-tunneling function, where the initial wave packet locations are indicated by gray circles at $(3N/4,5N/4+1)=(15,26)$.  The left-hand panel shows a moderate intensity enhancement below the threshold, $\gamma/\gamma_c=0.99$. The right-hand panel shows a dramatic intensity enhancement above the threshold, $\gamma/\gamma_c=1.01$; note the order-of-magnitude difference in the intensity scale. This order-of-magnitude enhancement occurs because the dimer eigenstates, with energies near the band edges, have a strong weight at 
the gain impurity site. 

These results show that due to the localized nature of all eigenstates, the time-dependent intensity profile $I(k,t)$ is acutely sensitive to parameters, such as $\beta$ and the initial state, which control whether the wave packet ``comes across'' the loss and gain impurities. Thus, we predict that {\it lattices with localized eigenstates provide a unique control over the violation of unitarity} - how much and how rapidly does the intensity change from its value in the Hermitian limit - that has no counterpart in traditional lattices with extended eigenstates. 

\section{Discussion}
\label{sec:disc}

In this paper, we have investigated the dynamics, disorder effects, and $\mP\mT$-symmetry breaking signatures in waveguide lattices that have, {\it in the disorder-free limit}, a majority of localized eigenstates. We have presented three, novel tunneling profiles, Eqs.~(\ref{eq:power})-(\ref{eq:ext}), that lead to such eigenstates. We have shown that the spatial spread of the wave packet and the frequency of Rabi oscillations can be controlled by the choice of the tunneling function and the initial position of the wave packet. We have also shown that the effect of weak disorder on such lattices is qualitatively different, and argued that, due to the localized nature of disorder-free lattice eigenstates, the signatures of $\mP\mT$-symmetry breaking in these lattices are acutely sensitive to the initial form of the wave packet. 

In this work, we have ignored the quartic interaction term that, in the case of optical waveguides, arises from nonlinear susceptibility and in the continuum limit, gives rise to the 
nonlinear Schr\"{o}dinger equation; this approximation is justified at low intensities. Since the interplay between interactions and disorder has profound effects on the phenomenon of localization~\cite{berg1,mit}, it will be interesting to explore them in the present system. 

Here, we have only focused on $\mP\mT$-symmetry breaking in even lattices with nearest neighbor impurities. Although Figs.~\ref{fig:far} and~\ref{fig:close} show the remarkable dependence of $\mP\mT$-symmetry breaking signatures on the initial wave packet, $\mP\mT$-symmetry breaking in even and odd lattices and the differences between them~\cite{derek} is an open question.  

Our results are not dependent on the exact forms of the tunneling functions, Eqs.~(\ref{eq:power})-(\ref{eq:ext}). Indeed, for example, any power-law tunneling function $(k^\mu,k^\nu)$ instead of $(k,k^2)$ will give similar results if $\mu\neq\nu$. Similarly, the $k-\beta\ln(k)$ tunneling function leads to a linear spectrum over a wide range $\beta$ because $\ln(x)$ grows more slowly than any power of $x$. The robustness of these results implies that the small, ubiquitous variations in the on-site potential and tunneling amplitudes in experimental samples will not affect our predictions. The experimental exploration of such coupled optical waveguides will deepen our understanding of lattice models with tunneling profiles whose continuum limit is different from the traditional Schr\"{o}dinger equation for a massive particle. 

\section*{Acknowledgments} 

This work was supported by the D.J. Angus-Scientech Educational Foundation (H.V.,V.V.,T.B.) and NSF Grant No. DMR-1054020 (Y.J.)

\end{document}